# Eclipsing Binaries in the Old Open Cluster NGC 2243

Janusz Kaluzny[1]

*Warsaw University Observatory, Al. Ujazdowskie 4, 00-478 Warsaw, Poland*

Wojciech Krzemiński

*Las Campanas Observatory, Casilla 601, La Serena, Chile*

Beata Mazur[1]

*Copernicus Astronomical Center, Bartycka 18, 00-716 Warsaw, Poland*

**Abstract.** We report the discovery of two contact binaries and two detached eclipsing binaries in the central part of NGC 2243.

NGC 2243 is an old open cluster located at a relatively high galactic latitude ($b = -18°$). Its color-magnitude diagram shows a clearly marked sequence of binary stars (Bonifazi *et al.* 1990, Bergbusch *et al.* 1991). These properties make it an attractive target in which to search for eclipsing binaries. The central part of NGC 2243 was surveyed for variable stars with the 1-m Swope telescope at Las Campanas Observatory. The observational data were collected during five runs conducted between December 1990 and December 1993. The area monitored ranged from $11 \times 11$ $(')^2$ to $21 \times 21$ $(')^2$ depending on the particular CCD camera used during a given run. We obtained a total of 367 frames in the V-band and 197 frames in the I-band. Exposure times ranged from 3 to 7 minutes. Four eclipsing binaries and one background RR Lyr star were discovered. Some basic data about these variables are given in Table 1. Two "comparison" fields located near the cluster were monitored for about 7 hours each. No variables were identified in either of these fields. This strongly suggests that both contact binaries discovered in the cluster field are members of NGC 2243. In Fig. 1 we show the location of the newly discovered variables on the cluster CMD. The light curve of variable V1 is presented in Fig. 2. This variable is a detached binary with a mass ratio close to unity and it is located near the turnoff of the cluster. A determination of the parameters of the components of V1 can provide direct information about the properties of turnoff stars in NGC 2243.

## References

Bergbusch, P.A., VandenBerg, D.A., & Infante, L. 1991, AJ, 101, 2102
Bonifazi, A., Fusi Pecci, F., Romeo, G., and Tosi, M. 1990, MNRAS, 245, 15

---

[1]Visiting Astronomer, Las Campanas Observatory

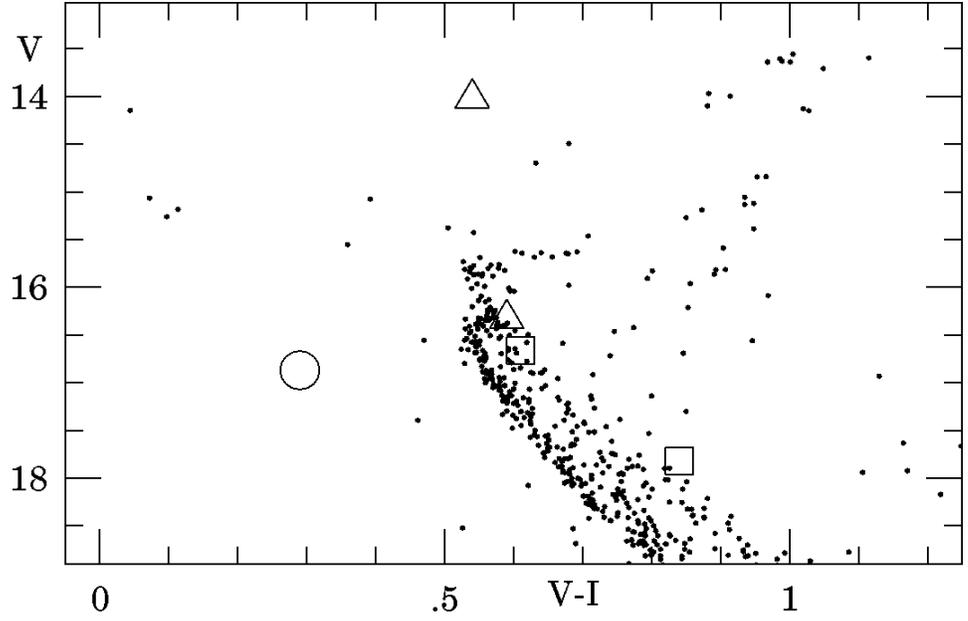

Figure 1. The CMD for the central part of NGC 2243 ($r < 3.2'$) with marked positions of five newly discovered variables. Triangles – EA systems; squares – contact binaries ; circle – RR Lyr star.

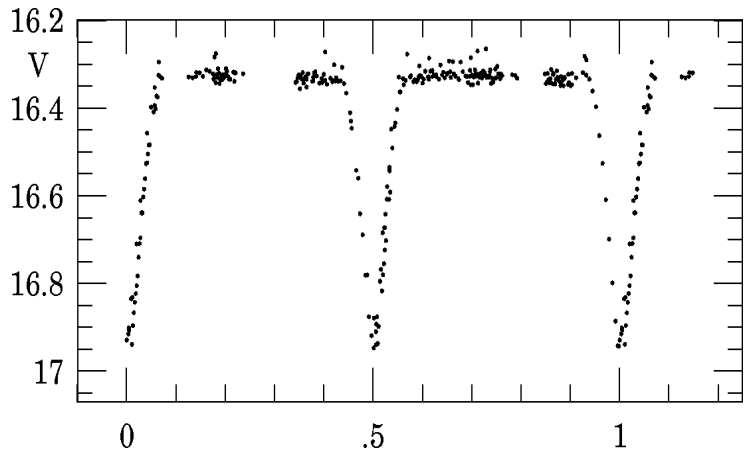

Figure 2. Phased light curve of variable V1

Table 1.    Variable stars discovered in the central part of NGC 2243

| ID | Type | Period days | Vmax | V-I |
|----|------|-------------|------|-----|
| V1 | EA | 1.188 | 16.33 | 0.59 |
| V2 | EW | 0.285 | 17.82 | 0.84 |
| V3 | EW | 0.356 | 16.66 | 0.61 |
| V4 | EA | ? | 14.02 | 0.54 |
| V6 | RR Lyr | 0.586 | 16.87 | 0.29 |